\documentclass[12pt]{article}
\usepackage{graphicx}

\newsavebox{\sboxpubnumber}
\newsavebox{\sboxpubdate}

\newcommand{\pubnumber}[1]{\begin{lrbox}{\sboxpubnumber}{\begin{tabular}{l} #1 \\
				 \usebox{\sboxpubdate}
				 \end{tabular}}
                           \end{lrbox}
                           \pubblock}
\newcommand{\Title}[1]{\begin{center} {\Large #1} \end{center}}
\newcommand{\Author}[1]{\begin{center} {\sc #1} \end{center}}
\newcommand{\Address}[1]{\begin{center} {\it #1} \end{center}}

\newcommand{\pubblock}{\rightline{
			\usebox{\sboxpubnumber}}}
\newenvironment{Abstract}{\begin{quotation}  }{\end{quotation}}
\newenvironment{Presented}{\begin{quotation} \begin{center}
             PRESENTED AT\end{center}\bigskip
      \begin{center}\begin{large}}{\end{large}\end{center}
      \end{quotation}}
\newcommand{\Acknowledgements}{\bigskip  \bigskip \begin{center} \begin{large}
             \bf ACKNOWLEDGEMENTS \end{large}\end{center}}
\def\ev{\,{\rm eV}}
\def\gev{\,{\rm GeV}}
\def\lessim{\mathrel{\rlap{\raise.5ex\hbox{$<$}}{\lower.5ex\hbox{$\sim$}}}}
\def\gtrsim{\mathrel{\rlap{\raise.5ex\hbox{$>$}}{\lower.5ex\hbox{$\sim$}}}}
\def\rmd{{\rm d}}

\def\GZK{{\rm GZK}}

\def\beq{\begin{equation}}
\def\eeq#1{\label{#1}\end{equation}}
\def\eeqn{\end{equation}}

\def\beqa{\begin{eqnarray}}
\def\eeqa#1{\label{#1}\end{eqnarray}}
\def\eeqan{\end{eqnarray}}

\let\bar=\overbar

\def\etal{{\it et al.}}

\def\Dslash{\not{\hbox{\kern-4pt $D$}}}
\def\dslash{\not{\hbox{\kern-2pt $\del$}}}

\def\msb{{\bar{\ssstyle M \kern -1pt S}}}

\begin{document}

%%%%%%%%%%%%%%%%%%%%%%%%%%%%%%%%%%%%%%%%%%%%%%%%%%%%%%%%%%%%%%%%%%%%%%%%
%%
%% START EDITING HERE!
%%
%%%%%%%%%%%%%%%%%%%%%%%%%%%%%%%%%%%%%%%%%%%%%%%%%%%%%%%%%%%%%%%%%%%%%%%%
\begin{titlepage}
%\pubdate{\today}          %fill in the date
\pubnumber{hep-ph/0202013} %preprint number(s)

\vfill
\Title{Ultra-high energy cosmic rays and new physics}
\vfill
\Author{Subir Sarkar}%\footnote{\tt s.sarkar@physics.ox.ac.uk}}
\Address{Department of Physics, University of Oxford,\\
         1 Keble Road, Oxford OX1 3NP, UK\\ E-mail: s.sarkar@physics.ox.ac.uk}
\vfill
%\andauth
%\vfill
%\Author{Your Coauthors Name}
%\Address{Department, Institute \\
%         Postal address}
\vfill
\begin{Abstract}
Cosmic rays with energies beyond the Greisen-Zatsepin-Kuzmin `cutoff'
at $\sim4\times10^{10}\gev$ pose a conundrum, the solution of which
requires either drastic revision of our astrophysical understanding,
or new physics beyond the Standard Model. Nucleons of such energies
must originate within the local supercluster in order to avoid
excessive energy losses through photopion production on the cosmic
microwave background. However they do not point back towards possible
nearby sources, e.g. the active galaxy Cen~A or M87 in the Virgo
cluster, so such an astrophysical origin requires intergalactic
magnetic fields to be a hundred times stronger than previously
believed, in order to isotropise their arrival
directions. Alternatively the primaries may be high energy neutrinos,
say from distant gamma-ray bursts, which annihilate on the local relic
background neutrinos to create ``Z-bursts''. A related possibility is
that the primary neutinos may initiate the observed air showers
directly if their interaction cross-sections are boosted to hadronic
strength through non-perturbative physics such as TeV-scale quantum
gravity. Or the primaries may instead be new strongly interacting
neutral particles with a longer mean free path than nucleons, coming
perhaps from distant BL-Lac objects or FR-II radio galaxies. Yet
another possibility is that Lorentz invariance is violated at high
energies thus suppressing the energy loss processes altogether. The
idea that has perhaps been studied in most detail is that such cosmic
rays originate from the decays of massive relic particles
(``wimpzillas'') clustered as dark matter in the galactic halo. All
these hypotheses will soon be critically tested by the Pierre Auger
Observatory, presently under construction in Argentina, and by
proposed satellite experiments such as EUSO.
\end{Abstract}
\vfill
\begin{Presented}
    COSMO-01 \\
    Rovaniemi, Finland, \\
    August 29 -- September 4, 2001
\end{Presented}
\vfill
\end{titlepage}
\def\thefootnote{\fnsymbol{footnote}}
\setcounter{footnote}{0}

%%%%%%%%%%%%%%%%%%%%%%%%%%%%%%%%%%%%%%%%%%%%%%%%%%%%%%%%%%%%%%%%%%%%%%%%
% The document starts here
%%%%%%%%%%%%%%%%%%%%%%%%%%%%%%%%%%%%%%%%%%%%%%%%%%%%%%%%%%%%%%%%%%%%%%%%
\section{Introduction}

In 1962, Linsley \cite{vr} detected a cosmic ray air shower at the
pioneering Volcano Ranch array with an estimated energy exceeding
$10^{11}\gev$. The significance of this event became clear only some
years later, following the serendipitous discovery of the universal
2.7 K cosmic microwave background (CMB), when Greisen and,
independently, Zatsepin and Kuzmin \cite{gzk} noted that cosmic ray
nucleons with energies exceeding $E_\GZK \sim 4\times10^{10}\gev$
would suffer severe energy losses through scattering on the CMB
photons due to excitation of the $\Delta\,(1232)$ resonance with a
cross-section of $\sim 400~\mu$b. Consequently their typical range
should decrease rapidly above this energy, leading to a `GZK cutoff'
in the energy spectrum if the sources are cosmologically
distant. Conversely the absence of such a cutoff in the spectrum
(which is approximated by a power-law $\rmd N/\rmd E \propto E^{-3}$
at lower energies) would imply that the sources are nearby, within the
local supercluster.\footnote{Recent detailed calculations
\cite{propag} indicate that the typical range of a nucleon drops below
$\sim100$~Mpc above $10^{11}\gev$. For heavy nuclei the dominant
energy loss process is photodissociation; above $10^{11}\gev$ this
happens through scattering on the CMB and the typical range is even
smaller than for nucleons \cite{nuclei}. For photon primaries the
dominant opacity above $E_\GZK$ is due to pair production on the
(poorly known \cite{radio}) extragalactic radio background; the mean
free path is estimated to be $\sim1-5$~Mpc at $10^{11}\gev$
\cite{photon}.}

Plausible astrophysical sources for such ultra-high energy cosmic rays
(UHECRs) were discussed by Hillas \cite{hil} who noted that simply to
contain charged particles long enough for them to be accelerated to
such high energies required sufficiently large magnetic fields and/or
spatial volumes such as could plausibly be achieved only in a few
specific sites such as the extended lobes of giant radio galaxies or
active galactic nuclei. Moreover particles of such energies should be
undeflected by the weak intergalactic magnetic fields, so should point
back to their sources. Thus there was the exciting prospect that
further such observations would clarify the long-standing question of
the origin of cosmic rays. Indeed UHECRs continued to be observed with
other detectors, notably at Haverah Park \cite{hp} and Yakutsk
\cite{yak}. However, perhaps because of the theoretical prejudice that
the sources of UHECRs must be cosmologically distant, as well as
concerns regarding uncertainties in the energy measurements, these
observations appear not to have been taken seriously \cite{review}. As
late as 1993 it was stated that the GZK cutoff is indeed present in
the data, confirming the expectation that the sources of UHECRs are
distant FR-II radio galaxies \cite{rsb93}.

The breakthrough came the same year with the detection of an event of
energy $(3.2\pm0.9)\times10^{11}\gev$ by the innovative Fly's Eye
detector in Utah, which measured the fluorescence from the excited
N$_2$ molecules in the atmospheric shower and was thus able to
reliably determine the energy \cite{fe}. Moreover the shower was seen
to have its maximum at an atmospheric depth of
$815^{+45}_{-35}$~g\,cm$^{-2}$, consistent with a proton primary, but
significantly less than the expected value of $1075$~g\,cm$^{-2}$ for
a photon primary \cite{hvsv95}.\footnote{Further observations by Fly's
Eye \cite{fecomp} as well as its successor HiRES \cite{hirescomp}
indicated a gradual increase in the depth of maximum with energy,
suggesting a change in composition from heavy nuclei to nucleons.}
Subsequently the Akeno Giant Air Shower Array (AGASA) has detected a
large number of air showers with energies exceeding $E_\GZK$
\cite{agasa}.

In fact the possible existence of such high energy cosmic rays had
already been mooted a decade earlier by particle physicists
\cite{shdm} as a likely signature of Grand Unified Theories (GUTs). In
particular, Hill \cite{td} had proposed that relic topological defects
such as magnetic monopoles created during GUT symmetry breaking in the
early universe may have formed metastable bound states which would
occasionally annihilate at the present epoch, thus releasing high
energy quark and gluon jets which would hadronise to release cosmic
rays of energy upto the GUT scale. Subsequently the development of
superstring theory had led to independently motivated suggestions for
the existence of massive relic particles associated with the `hidden
sector' of supersymmetry breaking; the lightest such state, having
only gravitational interactions, would be metastable, thus a natural
candidate for the dark mater \cite{crypton}. Indeed just before the
Fly's Eye event was announced, detailed studies had been undertaken of
the observational signatures of such particles \cite{us}, focussing on
the possibility of detecting high energy neutrinos from their slow
decays \cite{decay}. (Moreover, the possibility that weakly
interacting neutral particles such as neutrinos may acquire large
interaction cross-sections through new physics had been discussed in
the context of the detection of anomalous airshowers from the X-ray
binary Cygnus X-3 \cite{nustrong}).

In recent years several other speculative ideas have been put forward
for the possible origin of post-GZK UHECRs and some of them have been
studied in sufficient detail as to allow clear experimental tests. The
data set has also grown allowing some possibilities to be ruled out
altogether and severe constraints to be put on others. This is an
opportune time to assess the situation, particularly since the giant
Pierre Auger Observatory \cite{auger}, under construction in Argentina,
has recently seen ``first light'' and is expected to increase the
world statistics of UHECRs by ten-fold within a few years.

\section{Observational Status}

The AGASA experiment has provided most of the data on UHECRs in recent
years. In their 2000 review, Nagano and Watson \cite{review} had noted
that the measurements were consistent at the level of $\pm15\%$ in
energy (or $\pm45\%$ in flux) with older data from Haverah Park,
Yakutsk and Fly's Eye. After making various necessary corrections,
these authors had combined all the data to obtain the `standard'
differential energy spectrum:
\begin{eqnarray}
\label{standardspec}
 J(E) 
 =& C\left(\frac{E}{6.3\times10^{9}\gev}\right)^{-3.20\pm0.05},
 \quad &\mbox{for} \quad 4\times10^{8}\gev<E<6.3\times10^{9}\gev \\ \nonumber
 =& C\left(\frac{E}{6.3\times10^{9}\gev}\right)^{-2.75\pm0.20},
 \quad &\mbox{for} \quad 6.3\times10^{9}\gev<E<4\times10^{10}\gev,
\end{eqnarray} 
where
$C=(9.23\pm0.65)\times10^{-33}$m$^{-2}$s$^{-1}$sr$^{-1}$eV$^{-1}$.
Although the spectral shape at higher energies was uncertain, there
was no evidence of the GZK cutoff upto $\sim3\times10^{11}\gev$.
Preliminary data in 1999 from the HiRes air fluorescence detector
\cite{hires}, the successor to Fly's Eye, had 7 events above
$10^{11}\gev$, in conformity with this spectrum.

\begin{figure}[htb]
    \centering
    \includegraphics[height=3.4in]{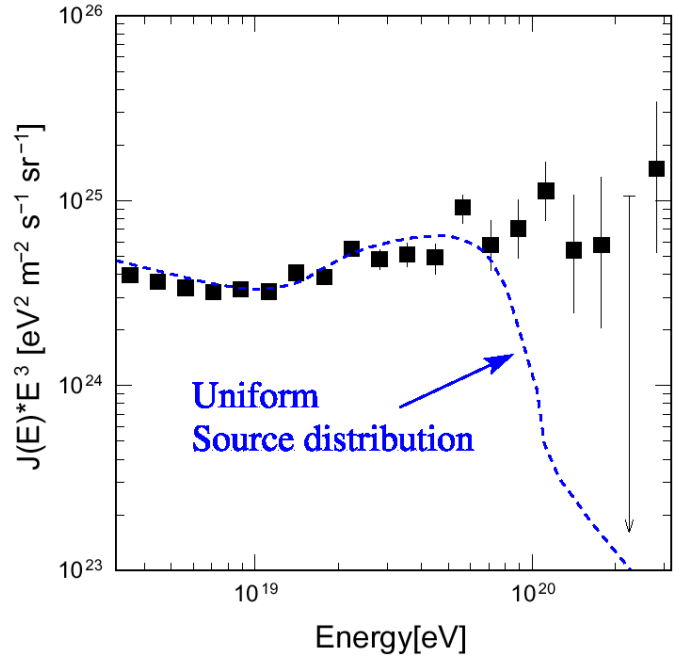}
    \caption{The AGASA spectrum compared with the expectation for the
    GZK-processed flux from cosmologically distant sources
    \protect\cite{agasanew}.}
    \label{agasanew}
\end{figure}

At the International Cosmic Ray Conference in Hamburg in August 2001,
AGASA reported 17 events above $10^{11}\gev$ with a total exposure of
$\sim6\times10^{16}$~m$^2$\,st\,s (the data set having been enlarged
by accepting showers inclined by as much as 60$^0$ to the vertical),
consistent with their earlier data \cite{agasanew}. On the basis of
more detailed comparison with simulations, the energy estimates were
raised (by 20\% at $10^{11}\gev$) relative to previously quoted
results. As shown in Fig.~\ref{agasanew}, the flux at $E>E_\GZK$ is
considerably above the expectation for an uniformly distributed
population of cosmologically distant sources (with a power-law
injection spectrum chosen to match data at lower energies).

\begin{figure}[htb]
    \centering
    \includegraphics[height=3.4in]{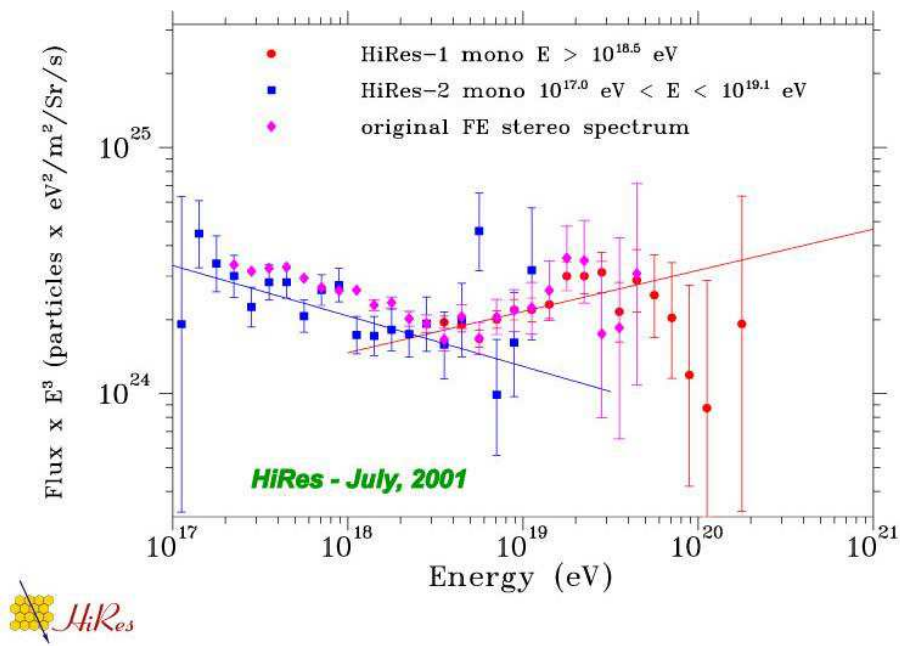}
    \caption{New results from HiRes \protect\cite{hiresnew} compared
    with the 2-component spectrum suggested by the older Fly's Eye
    data .}
    \label{hiresnew}
\end{figure}

However HIRES, with a similar exposure in the mono mode, reported only
2 events above $10^{11}\gev$ \cite{hiresnew}, instead of the $\sim20$
events expected on the basis of the above standard spectrum
(\ref{standardspec}). In stereo mode, with about 20\% of this
exposure, only 1 event was seen. As seen in Fig.~\ref{hiresnew}, there
is a clear discrepancy with the AGASA data. The resolution of this may
have to do with a possible underestimate of the efficiency of
atmospheric fluorescence emission at longer wavelengths (more relevant
for higher energies due to Rayleigh scattering) as suggested by a
recent measurement \cite{nagano}. This may have caused HiRes to
underestimate the energies (and overestimate the exposure) although
the magnitude of such corrections cannot yet be precisely quantified
\cite{watson}. It is also the case that the begining of the `ankle' in
the spectrum is a factor of $\sim2$ lower in the fluorescence detector
data than in the air shower detector data. The Auger Observatory
\cite{auger} employs both types of detectors so should be soon able to
resolve these discrepancies and establish a consistent energy
scale. However it should be emphasised that although the energy
spectrum (\ref{standardspec}) may require revision, the {\em
existence} of UHECRs above $E_\GZK$ is not in doubt!

\begin{figure}[htb]
    \centering
    \includegraphics[height=3.4in]{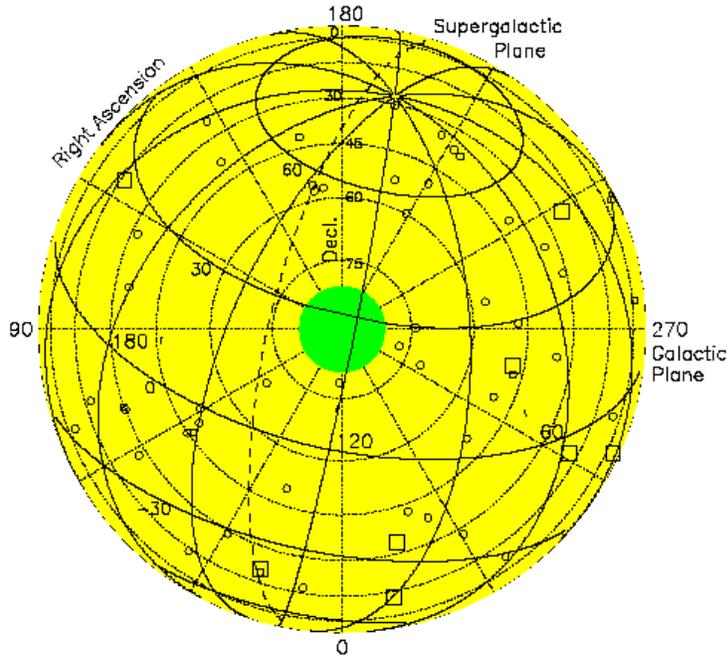}
    \caption{Arrival directions of UHECRs of energy
    $E>4\times10^{10}\gev$ (circles) and $E>10^{11}\gev$ (squares)
    \protect\cite{agasaanisonew}. The area is weighted according to
    the AGASA exposure (the shaded area is unobservable). }
    \label{agasaaniso}
\end{figure}

New results have also been presented by AGASA on the angular
distribution of events on the sky \cite{agasaanisonew}. In accordance
with earlier data, there is no evidence for any large-scale anisotropy
for the 59 observed showers (having zenith angles $<45^0$) with
energies above $E_\GZK$. As seen in Fig.~\ref{agasaaniso}, there is no
observed correlation with the galactic plane (such as is seen at lower
energies around $\sim10^9\gev$ by both Fly's Eye and AGASA
\cite{anisolow}), or with the super-galactic plane (as had been
claimed earlier, particularly in the Haverah park data
\cite{anisosuper}). However a number of `clusters' --- defined as a
grouping of 2 or more events within (approximately the experimental
angular resolution of) 2.5$^0$ --- are seen; the chance probability
that these result from an isotropic distribution is estimated by Monte
Carlo to be 0.05\% for the 5 observed doublets and 1.66\% for the 1
observed triplet \cite{agasaanisonew}. Assuming that {\em all}
observed events above $4\times10^{10}\gev$ come from one particular
class of source, the number of such sources is estimated assuming
Poisson statistics to be $220^{+207}_{-100}$. It is also found that
the cosmic rays contributing to the clusters have a hard energy
spectrum with a differential spectral index of $-1.8$. If these are
assumed to come from a separate source population then the total
number of observed sources is limited to be less than 427
\cite{agasaanisonew}. However it has been pointed out that the data
set used to make the initial claim for clustering ought not to be used
in the actual analysis; moreover the directions of the most energetic
events observed by previous experiments do not line up with any of the
6 `clusters' \cite{watson}. Thus the indication for compact sources is
exciting but still needs confirmation with increased statistics, as
will be provided by Auger.

\begin{figure}[htb]
    \centering
    \includegraphics[height=3in]{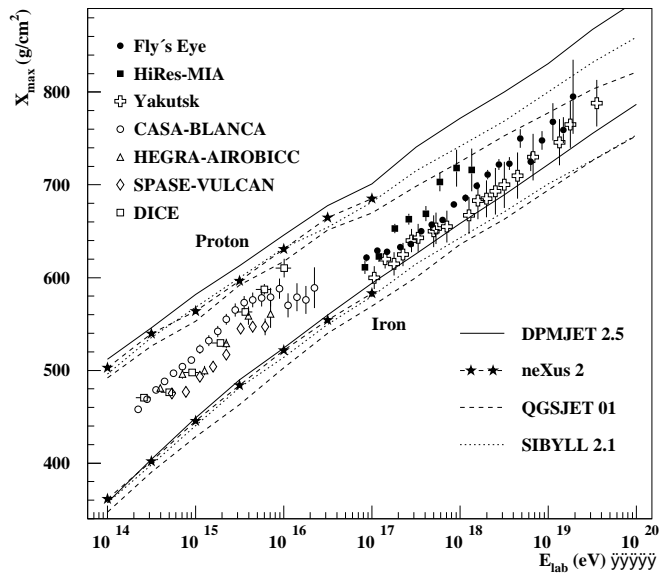}
    \caption{Variation of the depth of shower maximum with energy in
    different experiments, compared to expectations in various hadronic
    interaction models \protect\cite{comp}.}
    \label{xmax}
\end{figure}

Finally, with regard to the composition of UHECRs, new data has
muddled the picture suggested earlier by Fly's Eye and HiRes
\cite{fecomp,hirescomp} of a change from heavy nucleus (``iron'')
domination at $3\times10^8\gev$ to nucleon domination at
$10^{10}\gev$. The muon content of the AGASA showers implies a mixed
composition over $\sim10^9-10^{10}\gev$ \cite{agasacomp} and a new
analysis of Haverah Park data also indicates that nucleons make up no
more than 34\% of the flux in the range $(0.3-2)\times10^9\gev$
\cite{hpcomp}. As seen in Fig.\ref{xmax}, the conclusions are quite
sensitive to the choice of the interaction model used \cite{comp}
hence the matter cannot be considered to be definitively settled as
yet. However it would appear that a bound of $<50\%$ can be set on the
photon component at $E>E_\GZK$ from analysis of horizontal showers at
Haverah Park, independently of the interaction model \cite{hpcomp}. A
similar limit is obtained by AGASA on the basis that photon-initiated
showers tend to be muon-poor \cite{agasacomp2}. This is potentially a
powerful discriminant between models of the origin of
UHECRs. Unfortunately the propagation of photons at these energies is
uncertain since the dominant opacity (towards $\gamma\gamma$ pair
production) is determined by the low frequency radio background which
is poorly measured --- the last experimental attempt was in 1970
\cite{radio}. Subsequently model-dependent estimates have been made
which suggest a mean free path at $10^{11}\gev$ of a few Mpc
\cite{photon}. Hence even if photons are absent in the UHECRs observed
at Earth this does not exclude the possibility that they were emitted
by the sources, especially if the mean free path turns out to be even
smaller.

\section{Conventional explanations}

The lack of a GZK cutoff, which requires the sources to be relatively
nearby, coupled with the isotropy, which implies a cosmologically
distant population, pose severe problems for any astrophysical
expalanation of UHECRs \cite{blandford}. For example $\gamma$-ray
bursts are isotropically distributed and may well have the necessary
capability to accelerate protons to such high energies \cite{grb}, but
being cosmologically distant they {\em cannot} provide the events
observed at $E>E_\GZK$ \cite{nogrb}. Other possible sources such as
active galactic nuclei may even be able to account for the post-GZK
events if they are concentrated locally by a factor $\gtrsim30$
\cite{propag2}; however the local supercluster is overdense by a
factor of only $\sim2$. There are some nearby active galaxies but even
if they can accelerate UHECR these would need to be deflected through
large angles (and isotropised) by the intergalactic magnetic field
(IGMF) \cite{es95}; this requires the field strength to be $\sim100$
times stronger than the upper limit of a few nG inferred from
observational bounds on Faraday rotation in distant radio sources
\cite{igmf}. It has been proposed that the radio galaxy Cen~A at 3.4
Mpc is the source of UHECRs \cite{cena}; however detailed calculations
\cite{cenanot} show that this requires an IGMF of $\sim1\,\mu$G to
isotropise their directions and even so it would be difficult to
account for the direction of the highest energy Fly's Eye event. By
backtracing the 13 most energetic events in a magnetic field modelled
as originating from a `galactic wind', the radio galaxy M87 in the
Virgo cluster has also been implicated as the source \cite{m87}. The
convergence of the backtraced events turns out however to be due to
the {\em assumed} magnetic field geometry and cannot therefore be
considered evidence for an unique source \cite{m87not}.

\section{Models involving new physics}

Given the problems of an astrophysical origin, it is natural to
speculate that UHECRs are linked to physics beyond the Standard
Model. Many novel suggestions have been made in this regard
\cite{bsm}. I shall focus on only those which have been investigated
in sufficient detail so as to provide falsifiability criteria.

\subsection{Z-bursts}

Since at least one species of relic neutrinos must have a mass of
$\sim0.1\,\ev$, it is an attractive possibility to suppose that these
provide a target for ultra-high energy neutrinos from distant sources
to annihilate on. This would create ``Z-bursts'' with an energy of
$m_Z^2/2m_\nu\sim4\times10^{12}(m_\nu/1\,\ev)^{-1}\gev$, i.e. in just
the right energy range to be a source for UHECRs \cite{zburst}. The
energy spectrum of the nucleons and $\gamma$-rays resulting from $Z$
decays is well measured at LEP so a detailed comparison can be made
with the data \cite{fkr01}. In fact a good match to the energy
spectrum is obtained for a relic neutrino mass of
$0.26^{+0.2}_{-0.14}\,\ev$; however the required ultra-high energy
flux is rather large, taking into account that such light relic
neutrinos cannot have a overdensity within the GZK distance of more
than a factor of $\sim2$ \cite{fkr01}. Although there are no direct
experimental limits at such energies, any reasonable extrapolation to
somewhat lower energies would violate the limits obtained recently by
AMANDA. Moreover it is not clear how such high energy neutrinos would
be created in the hypothetical cosmic sources --- usually this would
require the acceleration of even higher energy nucleons! (In
principle, high energy neutrinos may be produced through the decays of
super-massive relic particles \cite{gk00} but such decays ought to
also produce UHECRs {\em directly} as we discuss below, making this
mechanism of less interest.) An alternative solution is to increase
the local density of relic neutrinos by postulating that they are
degenerate \cite{gk99}; however this is increasingly constrained by
failure to observed the associated `late ISW effect' in the angular
power spectrum of the CMB, as discussed by Hannestad at this meeting
\cite{steen}.

\subsection{Strongly interacting neutrinos}

Neutrinos with electroweak interactions cannot initiate the observed
airshowers but, as has been speculated for some time \cite{nustrong},
they may be become strongly interacting through new interactions. This
possibility has been recently revived with the realisation that the
string scale may in principle be as low as the TeV scale if there are
new dimensions in Nature. However as discussed by Pl\"umacher at this
meeting, the increase in cross-section through (t-channel) exchange of
Kaluza-Klein gravitons is inadequate to account for the
characteristics of the observed events \cite{nunot}. A
string-theoretic calculation in which the string amplitude is
approximated as a sum over s-channel resonances (leptoquarks) also
comes to a negative conclusion \cite{stringnu}.

\subsection{New neutral primaries}

The GZK cutoff would be higher if the primary particle is a new stable
hadron which is heavier than a nucleon. For example it might be a
bound state of a light gluino if that is the lightest supersymmetric
particle \cite{gluino} or perhaps a H-dibaryon \cite{kochelev}. The
air showers that would be initiated by such a particle (``uhecron'')
have been studied in detail \cite{uhecron} and comparison with data
limits the primary mass to be $\lessim50\gev$. The experimental
constraints on new stable hadrons are already quite stringent
\cite{pdg} and it is likely that such a particle can soon be found or
excluded. The interactions needed to produce such particles in distant
sources must also produce neutrinos and $\gamma$-rays and the latter
should be detectable, e.g. by C\^erenkov telescopes such as VERITAS,
MAGIC and HESS \cite{bsm}.

\subsection{Violation of Lorentz invariance}

It has been noted that a small modification of the relation between
momentum and energy in special relativity may undo the GZK cutoff
\cite{lv}. In particular this may happen for a deformed dispersion
relation $c^2p^2=E^2(1+E/M_{\rm P})+\ldots$, motivated by
considerations of quantum gravity at the Planck scale
\cite{gac}. However this needs further examination in a quantum
gravitational framework before definite conclusions can be drawn
since, e.g., energy might be conserved only in a statistical sense
\cite{ellis}. 

\subsection{Decaying supermassive dark matter}

As mentioned earlier, the possible existence of relic metastable
massive particles whose decays can create high energy cosmic rays and
neutrinos had been discussed \cite{shdm,crypton,us,decay} {\em before}
the famous Fly's Eye event \cite{fe} which focussed attention on the
enigma of UHECRs. Subsequently this idea was revived \cite{kr98} and
it was further pointed out that such particles, being cold dark
matter, would naturally have a overdensity by a factor of $\sim10^4$
in the halo of our Galaxy \cite{bkv97,bs98}. Hence if their slow
decays generate UHECRs, all propagation effects will be unimportant
(except possibly for photons) in determining the observed spectrum and
cosmposition, and, moreover, there should be a detectable anisotropy
\cite{aniso} in the arrival directions, given our asymmetric position
in the Galaxy. In order to account for the highest energy events with
the observed rates, the particle mass should exceed
$m_X\gtrsim10^{12}\gev$ while its lifetime is determined to be
$\tau_X\simeq3\times10^9\zeta_X\,t_0$, where $t_0\sim10^{10}$~yr is the
age of the universe and $\zeta_X$ is the fraction of the halo dark
matter density in the form of such particles \cite{cosmo99}. This is
in accordance with the theoretical expectation for `cryptons'
\cite{crypton}. Moreover, as discussed by Kolb at this meeting, it is
plausible that such particles (``wimpzillas'') were created with a
cosmologically interesting abundance through the changing
gravitational field at the end of inflation \cite{wimpzilla}.

\begin{figure}[htb]
    \centering
    \includegraphics[height=3in]{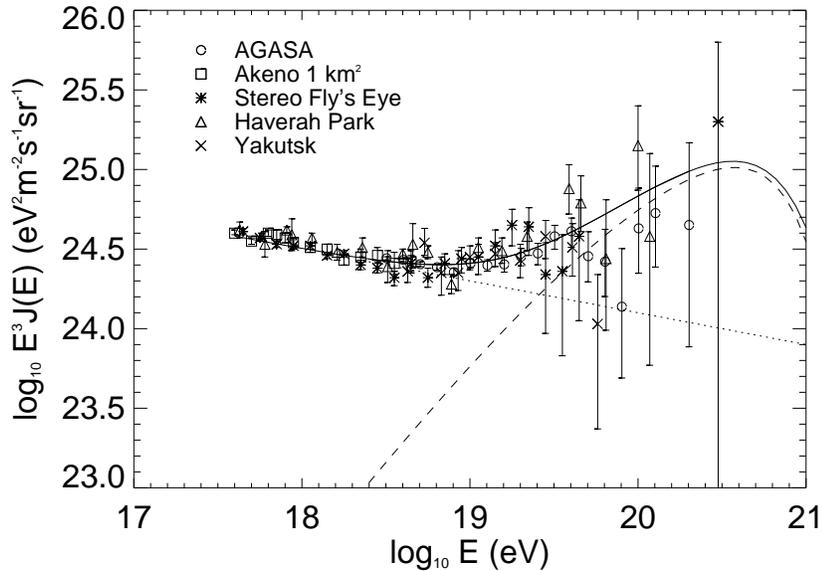}
    \caption{Fit to the UHECR spectrum beyond the `ankle' with a
     decaying dark matter particle mass of $5\times10^{12}\gev$
     (dashed line), including the effects of supersymmetry 
     \protect\cite{st02}.}
    \label{frag}
\end{figure}

The spectra of the decay products is essentially determined by the
physics of QCD fragmentation \cite{cosmo99}. Recently several groups
have undertaken calculations which improve on previous efforts
\cite{td,bkv97,bs98} in being both better connected to laboratory data
(available upto LEP energies) and incorporating the possible effects
of new physics e.g. supersymmetry at higher energies
\cite{rubin,fk01,zia01,bk01,cf01,st02}. As discussed by Toldra
\cite{toldra} at this meeting, an effective approach is to evolve the
fragmentation functions measured at the $Z^0$ peak (for neutrinos,
inferred using HERWIG) upto the scale of the decaying particle mass by
solution of the DGLAP evolution equations. Similar results are
obtained in two independent calculations \cite{fk01,st02} and, as seen
in Fig.~\ref{frag}, the evolved spectrum of nucleons is in good
agreement with the `flat' component (\ref{standardspec}) of cosmic
rays at $E>E_\GZK$. On the negative side, the decay photons, which
have a similar spectral shape are more abundant by a factor of
$\sim2$, so this model (as well as the related one involving
annihilations of wimpzillas \cite{ann}) would seem to be ruled out by
the Haverah Park bound \cite{hpcomp} on the photon component of
UHECRs. However, as commented earlier, it is quite possible that such
high energy photons are significantly attenuated in their passage
through the low frequency radio background; this would in turn
generate a background of low energy $\gamma$-rays and it is necessary
to check that this does not exceed observational limits (work in progress).

\begin{figure}[htb]
    \centering
    \includegraphics[height=3.3in]{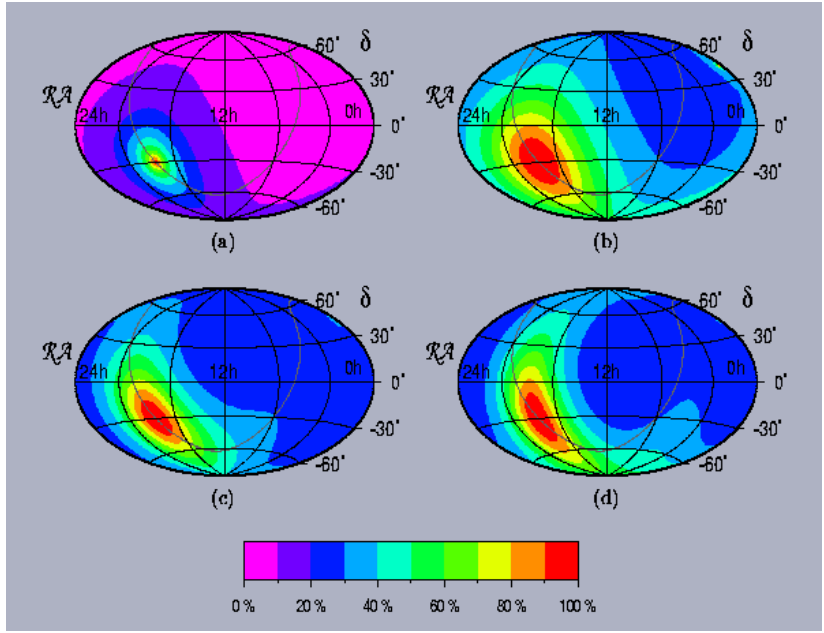}
    \caption{Contour plots (Hamer-Aitoff projections) of the UHECR sky
     for (a) cusped, (b) isothermal, (c) triaxial and (d) tilted
     models of the dark matter halo of our Galaxy. The effect of the
     halo of M31 is visible in the upper right of each
     plot \protect\cite{efs01}.}
    \label{halo}
\end{figure}

Detailed calculations have also been made of the expected anisotropy,
adopting different possible models of the dark matter halo (cusped,
isothermal, triaxial and tilted) \cite{efs01}. We find that the
amplitude of the anisotropy is controlled by the extent of the halo,
while the phase is controlled by its shape. As seen in
Fig.~\ref{halo}, the amplitude of the first harmonic is $\sim0.5$ for
a cusped halo, but falls to $\sim0.3$ for an isothermal halo, while
the maximum is in the direction of the Galactic Centre, with
deviations of up to $30^0$ possible for triaxial and tilted
haloes. Contrary to a recent claim \cite{arnold} the halo of M31 is
not bright enough to provide conclusive evidence for this hypothesis.
However another key signature of a dark matter origin may be some
`clustering' of events since the halo CDM distribution is expected to
be clumpy rather than being smooth \cite{clustdm}. Thus this model has
several interesting observational signatures and can easily be
falsified, e.g. if the UHECRs turn out to be heavy nuclei \cite{heavy}. 

\section{Conclusions}
After a long hiatus, high energy cosmic rays have again become very
interesting for particle physicists looking for evidence of physics
beyond the Standard Model. The source of the highest energy particles
in Nature is an equally interesting enigma for astrophysicists. As
Lema\^{\i}tre first suggested, the origin of such particles may even
be linked to the early universe, although not quite as he imagined!
Presently the data are tantalising but not sufficient in either
quantity or quality to distinguish definitively between proposed
models. The good news that this will soon be remedied by the Auger
array \cite{auger} which should be complete by 2004 and is expected to
detect $\sim2000$ events above $E_\GZK$ within 5 years. Moreover the
ambitious space-based experiment EUSO \cite{euso}, scheduled for a
3-year flight on the International Space Station in 2007, will provide
another substantial increase in collecting power. This is an exciting
time for cosmic ray physics and we look forward to the surprises that
Nature has in store for us.

\Acknowledgements I would like to thank Matts Roos and the other
organisers of this stimulating meeting for the invitation to speak,
and Wyn Evans, Francesc Ferrer and Ramon Toldra for their
collaboration in the work presented here. I am grateful to Alan Watson
for communicating his expert views on the experimental situation.

\end{document}